\documentclass[aps,preprint,showpacs,preprintnumbers,amsmath,amssymb,prd]{revtex4}
\pdfoutput=1
\usepackage{amssymb}
\usepackage{graphicx}

\usepackage{graphics,psfrag}
\usepackage{graphicx,psfrag}
\usepackage{epic,eepic}
\usepackage{color}
\usepackage{epsfig}
\usepackage{latexsym}
\usepackage{pstricks}

\def\be{\begin{equation}}
\def\ee{\end{equation}}
\def\ba{\begin{eqnarray}}
\def\ea{\end{eqnarray}}

\begin{document}

\title{Bending of electromagnetic wave in an ultra-strong magnetic field }

\author{Jin Young Kim }
\email{jykim@kunsan.ac.kr}

\affiliation{Department of Physics, Kunsan National University,
Kunsan 573-701, Korea}

%\date{\today}

\begin{abstract}
We consider the bending of light by nonlinear electrodynamics when
the magnetic field $B$ exceeds the critical value $B_{\rm c} = m^2
c^2 /e \hbar = 4.4 \times 10^9 \rm T$. Using the index of refraction
derived from the analytic series representation in one-loop
effective action of QED, we found the trajectory and the bending
angle of light in geometric optics. The angle bent by ultra-strong
magnetic field of magnetar was estimated and compared with the
gravitational bending. The result may be useful in studying the
lensing, birefringence, and other nonlinear quantum electrodynamic
effects above  $B_{\rm c}$.

\end{abstract}

\pacs{12.20.Fv,41.20.Jb,95.30-k}

\maketitle

%\section{Introduction}

Electrodynamics is nonlinear not only in a continuous medium but
also in a non-trivial vacuum by quantum correction. The vacuum
polarization by virtual electron and positron in the presence of the
non-trivial quantum electrodynamics (QED) vacuum causes the velocity
shift and can be described by the index of refraction. For example,
electromagnetic field can induce a material-like behavior by
non-local features of the quantum vacuum structure. In the case of a
weak field, the so-called Euler-Heisenberg effective Lagrangian
\cite{eulhei,schwinger,Bialynicka,adler} is a result of this quantum
correction.

The fundamental scale for vacuum polarization is set by the critical
field of QED, $B_{\rm c} = E_{\rm c}/c =  m^2 c^2 / e \hbar = 4.4
\times 10^9 \rm T$. Nonlinear properties by non-trivial QED vacua
are being tested and confirmed with the advances in experimental
devices, especially in high power laser technology \cite{larmoreaux,
burkeetal,Piazza06,kpk,woojoong,sushkov}. However, as far as the
phenomena related to the strong electromagnetic field are concerned,
it is very difficult to test them in ground laboratories since the
maximum available field is of the order $B \sim 10^2 \rm T$. One can
think of the astrophysical objects as sources of the strong
electromagnetic field. A significant growth in this field of
research has been done due to the observation of strong magnetic
field in neutron star \cite{lorenci01,
shaviv,denisov,denisov01,heylshav02,denisov03,heylshav03,
denisov04,denisov05,dupays,denisov07,heyl10,kim1,kim2}.

The light bending by a massive object is a result of general
relativity and is a useful tool in astrophysics. The light bending
can also happen when the light ray passes electrically or
magnetically charged objects by the nonlinear interaction of QED. In
geometric optics, light path can be bent by a continually varying
index of refraction. If there is a gradient in the index of
refraction by non-trivial QED vacuum, one can calculate the bending
angle in analogy with the geometric optics. For example, Denisov
{\it et al.} \cite{denisov} calculated the bending angle using the
eikonal equation when a ray is passing through the equator of the
magnetic dipole. In the previous work of the author \cite{kim1}, the
trajectory of light and bending angle in a Coulombic field were
computed in geometric optics based on the Euler-Heisenberg effective
action. Also the bending angle of light by the electric field of a
charged black hole was estimated and a general formula of the
bending angle valid for any orientation of the magnetic dipole was
derived \cite{kim2}.

In astrophysics there are objects with very strong electric or
magnetic field. As far as the magnetic field is concerned, there are
neutron stars with the surface magnetic field above the QED critical
field, known as magnetars. Since the Euler-Heisenberg effective
action is represented as an asymptotic series, its application is
confined to the weak field limit. So the result based on the
Euler-Heisenberg Lagrangian is not valid in this ultra-strong field
limit. The knowledge on the effective action for strong field limit
is crucial to study the nonlinear electromagnetic properties above
the critical field $B_{\rm c}$. In this paper, we will study the
bending of light in the ultra-strong magnetic field of magnetars.

%\section{The index of refraction and trajectory equation}

To study the light bending above the critical field, we need the
complete expression of the action. Recently, analytic series
representation for the one-loop effective action of QED has been
studied from Schwinger's integral form
\cite{heyher97,heyher,ditgies,chopak01,chopak06,huliu}. When the
electric or magnetic field is above the QED critical limit, light
propagating in such ultra-strong electromagnetic fields no longer
satisfied the light cone condition, i.e., $v \ne c$. We use the
explicit expression for the average velocity of light in
ultra-strong electric and magnetic fields derived by Cho {\it et
al.} \cite{chopak06}
 \ba
 {\bar v}^2_{\rm m} &\simeq&
 \frac{ 1 - {e^2 \over {12 \pi^2} } ( \ln \frac{eB}{m^2} + c_1 +1 ) }
      { 1 - {e^2 \over {12 \pi^2} } ( \ln \frac{eB}{m^2} + c_2)
         + \frac {e^3 B}{24 \pi^2 m^2} } ,   \label{vsquaremag} \\
 {\bar v}^2_{\rm e}  &\simeq&
 \frac{ 1 - {e^2 \over {12 \pi^2} } ( \ln {eE \over m^2} + c_2) }
      { 1 - {e^2 \over {12 \pi^2} } ( \ln {eE \over m^2} + c_1 + 1)
      },       \label{vsquareel}
 \ea
where
 \ba
  c_1 &=& - \gamma - \ln \pi + {6 \over \pi^2} \zeta^\prime (2) =
  -2.29191\dots,     \nonumber   \\
 c_2 &=& \ln 2\pi + c_1 + {3 \over 2} - \gamma = -1.82070\dots .
   \nonumber
 \ea
 The corresponding index of refraction is given by
 $n=1/v$. Up to the leading order in one-loop,
 one can approximate the index of refraction $n$ by
  \ba
  {\bar n}_{\rm m} &\simeq& \sqrt{ 1 + \frac{e^2}{12 \pi^2} (c_1 -c_2 +1)
  + \frac {e^3 B}{24 \pi^2 m^2}  },  \label{nmag}   \\
  {\bar n}_{\rm e} &\simeq& \sqrt{1 - \frac{e^2}{12 \pi^2} (c_1 -c_2 +1) }.  \label{nelec}
  \ea

Now we consider the bending of light when it passes near the
astronomical object with strong electric or magnetic field above the
critical value. As a source of ultra-strong electric field one can
think of a charged black hole theoretically. As a source of
ultra-strong magnetic field one can consider the magnetar whose
magnetic field strength on the surface is estimated up the order of
$10^{11} \rm T$. The physical limit to the magnetic field of neutron
star is of the order $10^{12}-10^{14} \rm T$ \cite{zaumen,lerche}.
Beyond this limit, the fluid inside the star would mix and the
magnetic field would dissipate. No objects in the universe can
maintain fields stronger than this limit. Thus we consider the
maximal field strength up to $10^{12} \rm T$. Note that the index of
refraction is close to one for the range of our consideration (see
figure 1).

\begin{figure}
\includegraphics[angle=0, width=8cm ]{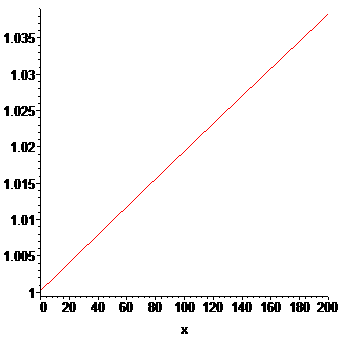} \caption{
Index of refraction ${\bar n}_{\rm m}$ plotted as a function of $x =
B/B_{\rm c}$ up to $x =200 (\sim 10^{12}/4.4 \times 10^9)$. }
\label{fig1}
\end{figure}

Eq. (\ref{nelec}) tells us that there is no significant change of
refraction index by ultra-strong electric field. So it is expected
that the bending of light is negligible when photons pass the
ultra-strong electric field of a charged black hole. To calculate
the bending angle by the ultra-strong magnetic field of magnetar, we
consider the trajectory of a light ray when it passes the magnetic
field of a magnetic dipole. In general the bending angle will depend
on both the orientation of the dipole relative to the direction of
the incoming ray and the polarization of the light. To simplify the
calculation, we consider the case when the photon path is
perpendicular to the dipole moment and traveling on the equator of
the dipole. The generalization for arbitrary orientation of the
dipole axis can be straightforward as in \cite{kim2}. We consider
the magnetic dipole located at the origin directing along the $z$
axis and the incident photon is coming from $x= -\infty$ with an
impact parameter $b$ (see figure 2).

\begin{figure}
\includegraphics[angle=0, width=8cm ]{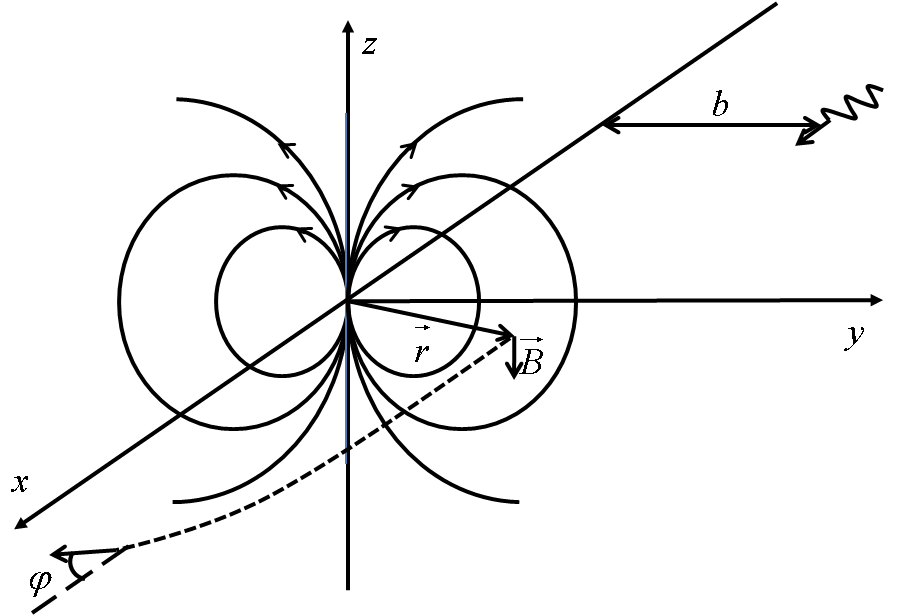} \caption{
Schematic of light bending by magnetic dipole when the photon path
is on the equator of the dipole.} \label{fig2}
\end{figure}

Because the quantum correction in the presence of ultra-strong
magnetic field is represented in terms of index of refraction, we
can consider the analogous light propagation in a classical medium
within geometric optics formalism. From the Snell's law, the
trajectory equation can be written as \cite{kim1,kim2}
 \be
 \frac{d{\bf u}}{ds}=\frac{1}{n}({\bf u}\times {\nabla}n)\times {\bf u},
 \ee
where ${\bf u}$ denotes the unit vector in the direction of
propagation, $s$ denotes the distance parameter
 of the light trajectory with $ds=|d\vec{\bf r}|$, and
 \be
 {\bf u}=\frac{d\vec{\bf r}}{ds}.
 \ee
 When the correction to the index of refraction is small,
 the trajectory equation can be approximated to the leading order as
 \be
 \frac{d{\bf u}}{ds}=({\bf u}_0\times {\nabla}n)\times {\bf u}_0,
 \ee
 where ${\bf u}_0$ denotes the initial direction of the incoming photon.
For a photon coming in from $x=-\infty$ and moves to $+x$ direction,
 \be
 {\bf u}_0=(1,0,0),
 \ee
 and defining ${\nabla}n \equiv (\eta_1,\eta_2,\eta_3)$ the trajectory equation becomes
 \be
 \frac{d^2x}{ds^2}=0, ~~~~ \frac{d^2y}{ds^2}=\eta_2, ~~~~
 \frac{d^2z}{ds^2}=\eta_3 .
 \ee
 The first equation shows that
$ds=dx$ at the leading order and the trajectory equations for $y(x)$
and $z(x)$ are given by
 \be
 \frac{d^2y}{dx^2}=\eta_2, ~~~~~~
 \frac{d^2z}{dx^2}=\eta_3,  \label{trajectory}
 \ee
 which will be used in the following analysis.
Note that, on the equator of the dipole, there is no gradient in $z$
direction by symmetry, $\eta_3 = 0$. Thus there is no bending in $z$
direction and the bending will occur toward the center of the
dipole.

%\section{Light bending by magnetar}

The magnetic field at a distance $r$ on the equator of the magnetic
dipole is given by
 \be
 {\rm B} = B_0 \frac{r_0^3}{r^3} (- \hat z ),
 \ee
where $B_0$ is the magnetic field on the surface of the neutron
star. Restoring the physical constants, the index of refraction can
be written as
 \be
  {\bar n}_{\rm m} \simeq \sqrt{ 1 + \frac{\alpha}{3 \pi} (c_1 -c_2 +1) + \frac{ \alpha}{6 \pi}
  \frac{B_0}{B_{\rm c}} \left( \frac{r_0}{r} \right)^3
  },  \label{nmagequator1}
  \ee
where $\alpha = 1/137$ is the fine structure constant. As shown in
figure 1, the index of refraction does not deviate much from one up
to $B_0 = 10^{12} \rm T$. Thus, we approximate the index of
refraction as
 \be
  {\bar n}_{\rm m} \simeq  1 +  \frac{\alpha}{6 \pi} (c_1 -c_2 +1)
  + \frac{ \alpha}{12 \pi}
  \frac{B_0}{B_{\rm c}} \left( \frac{r_0}{r} \right)^3
  ,  \label{nmagequator2}
  \ee
and the trajectory equation can be written explicitly as
 \be
 \frac{d^2y}{dx^2}= \eta_2 = - \frac{\alpha}{4\pi} \frac{B_0
 r_0^3}{B_{\rm c}} \frac{y}{r^5}.    \label{trajecdiffeq}
 \ee

For the incoming photon with the impact parameter $b$, the initial
condition reads
 \be
 y(-\infty) = b, ~~~~y^{\prime} (-\infty) = 0.
 \ee
 Integrating Eq. (\ref{trajecdiffeq}) by putting $y=b$
 for the leading order solution, we obtain
 \ba
 y^\prime (x) &=& - \frac{\alpha}{12\pi} \frac{B_0}{B_{\rm c}} \frac{r_0^3}{b^3}
 \left[ \frac{x}{\sqrt{b^2+x^2}} \left( \frac{b^2}{b^2 + x^2} +2 \right) +2
 \right],      \\
 y(x) &=& b \left[ 1 - \frac{\alpha}{12\pi} \frac{B_0}{B_{\rm c}} \frac{r_0^3}{b^3}
 \left( 2 \frac{\sqrt{b^2 + x^2} +x}{b} - \frac{b}{\sqrt{b^2+x^2}}
 \right) \right].
 \ea
The total bending angle $\varphi_{\rm m}$ can be obtained from
$y^\prime (\infty)$,
 \be
 | y^\prime (\infty) | = \tan \varphi_{\rm m} \simeq \varphi_{\rm m}
 ,
 \ee
and is given by
  \be
  \varphi_{\rm m} = \frac{\alpha}{3\pi} \frac{B_0}{B_{\rm c}}
  \frac{r_0^3}{b^3}.  \label{varphiav}
  \ee

Compared with the bending angle obtained from the asymptotic form of
Euler-Heisenberg effective action valid in the weak field limit
\cite{denisov,kim2},
 \be
 \varphi_{\rm mEH} \propto \frac{B_0^2}{b^6},
 \ee
the power dependence $\varphi_{\rm m} \propto {B_0}/{b^3}$ implies
that the magnetic bending may be important at short distance. Let us
estimate the bending angle for an ultra-strongly magnetized neutron
star and compare it with the gravitational bending. According to the
general relativity, the gravitational bending angle is given by
 \be
 \varphi_{\rm g} = \frac{4G {\cal M}}{bc^2}.   \label{gbending}
 \ee
Consider a neutron star with mass $ {\cal M}= {\cal M}_{\rm sun} = 2
\times 10^{30} {\rm kg}$ and radius $r_0 = 10 {\rm km}$.
Parameterizing the impact parameter in units of the radius $b =\zeta
r_0 $ with $\zeta
>1$, the bending by a ultra-strong magnetic field can be expredded as
 \be
 \varphi_{\rm m} = \frac{\alpha}{3\pi} \frac{B_0}{B_c} \frac{1}{\zeta^3}.
  \label{mbending}
 \ee
The maximal gravitational bending for $\zeta =1$ ($b= r_0$) is
$\varphi_{\rm g} = 0.59 {\rm rad}$. For a magnetar with the field
strength on the surface $B_0 = 10^{11} {\rm T}$, the maximal value
of the magnetic bending angle is estimated as $\varphi_{\rm m}  =
1.8 \times 10^{-2} {\rm  rad}$. This value is one order of magnitude
smaller than the gravitational bending. For $B_0 = 10^{12} {\rm T}$,
the magnetic bending can be comparable to the gravitational bending.

%\section{discussion}

So far we have considered the light bending by average value of the
index of refraction. In the presence of strong magnetic field, QED
renders the vacuum birefringent; the light velocities for
perpendicular and parallel modes are different. The asymptotic
formula for the light velocities for the ultra-strong field limit
are given by \cite{chopak06}
 \ba
 {v}_\bot^2 &\simeq&
 \frac{ 1 - {e^2 \over {12 \pi^2} } ( \ln {eB \over m^2} + c_1 + {3 \over 2} ) }
      { 1 - {e^2 \over {12 \pi^2} } ( \ln {eB \over m^2} + c_1
      + {1 \over 2}) } ,         \label{vsquareperp}        \\
 {v}_\|^2 &\simeq&
 \frac{ 1 - \frac{e^2}{12 \pi^2} ( \ln \frac{eB}{m^2} + c_1 + \frac{1}{2} ) }
      { 1 - \frac{e^2}{12 \pi^2} ( \ln \frac{2eB}{\pi m^2} + 1 - \gamma )
      + \frac{e^3 B}{12 \pi^2 m^2} } .   \label{vsquarepar}
 \ea
The corresponding leading order terms of the indices of refraction
are
  \ba
  {n}_{\perp} &\simeq& 1+ \frac{e^2}{24 \pi^2},   \label{nperp} \\
  {n}_{\parallel} &\simeq& 1 + \frac{e^2}{24 \pi^2} (c_1 - \frac{1}{2}
  +\ln \frac{\pi}{2} + \gamma)
  + \frac {e^3 B}{24 \pi^2 m^2}  .   \label{npara}
  \ea
Compared with Eq. (\ref{nmag}), the bending angle for the parallel
mode will be twice the value given by Eq. (\ref{varphiav}) while
there is no bending for the perpendicular mode. This birefringence
effect can be one way of testing the validity of our calculation if
the allowed precision in measuring the photon polarization is
enough.

Note that, from Eqs. (\ref{varphiav}) and (\ref{gbending}), the
relative bending $ \varphi_{\rm m} / \varphi_{\rm g} \propto 1/b^2$.
For example, at impact parameter $b= 10 r_0$, the magnetic bending
$\varphi_{\rm m}  = 1.8 \times 10^{-5} {\rm  rad}$ is three orders
of magnitude smaller then the gravitational bending $\varphi_{\rm g}
= 5.9 \times 10^{-2} {\rm rad}$. Even if the magnetic bending angle
may be small compared with the gravitational bending, one can check
the result in this work by measuring the total bending angles for
different values of the impact parameter. Since the power dependence
of two bending angles on the impact parameter is different,
$\varphi_{\rm g} \propto 1 /b, ~\varphi_{\rm m} \propto 1/b^3 $, one
may test the power dependence by fitting the data from different
impact parameters to $\varphi_{\rm tot} = \varphi_{\rm g} +
\varphi_{\rm m}$.

The constraint on strong B-field consistent with the one-loop
approximation is $B/B_{\rm c} < \pi/ \alpha \simeq 430$. The maximum
field strength that we considered in this work, $ B \sim 10^{12} \rm
T$ ($x = B/B_{\rm c} \simeq 200$), is within this range. So our
result may be useful in studying the lensing, birefringence, and
other nonlinear quantum electrodynamic effects by ultra-strong
magnetic field of magnetars.

\acknowledgements{ We would like to thank M. I. Park  for discussion
and help. This research was supported by Basic Science Research
Program through the National Research Foundation of Korea (NRF)
funded by the Ministry of Education, Science and Technology
(12A12840581). }

\end{document}